\documentclass[]{proarticle}
\usepackage{multicol}
\usepackage{graphicx}
\usepackage{amssymb}
\usepackage{epsfig}

\def\bea{\begin{eqnarray}}
\def\eea{\end{eqnarray}}
\begin{document}

\title{Cosmographic transition redshift in $f(R)$ gravity}

\author{Salvatore Capozziello${}^{1,2,3}$ and Orlando Luongo${}^{1,2,4}$}

\maketitle

\address{${}^{1}$Dipartimento di Fisica, Universit\`a di Napoli ``Federico II", Via Cinthia, I-80126, Napoli, Italy.}
\address{${}^{2}$Istituto Nazionale di Fisica Nucleare (INFN), Sez.\ di Napoli, Via Cinthia, I-80126, Napoli, Italy.}
\address{${}^{3}$Gran Sasso Science Institute (INFN), Viale F. Crispi, I-67100, L'Aquila, Italy.}
\address{${}^{4}$Instituto de Ciencias Nucleares, Universidad Nacional Aut\'onoma de M\'exico, DF 04510, M\'exico, Mexico.}

\eads{capozzie@na.infn.it, luongo@na.infn.it}


\begin{abstract}
We propose a strategy to infer the transition redshift $z_{da}$, which characterizes the passage through the universe decelerated to accelerated phases, in the framework $f(R)$ gravities. To this end, we numerically reconstruct $f(z)$, i.e. the corresponding $f(R)$ function re-expressed in terms of the redshift $z$ and we show how to match $f(z)$ with cosmography. In particular, we relate $f(z)$ and its derivatives to the cosmographic coefficients, i.e. $H_0, q_0$ and $j_0$ and  demonstrate that its corresponding evolution may be framed by means of an effective logarithmic dark energy term $\Omega_X$, slightly departing from the case of a pure cosmological constant. Afterwards, we show that our model predicts viable transition redshift constraints, which agree with  $\Lambda$CDM. To do so, we compute the corresponding $z_{da}$ in terms of cosmographic outcomes and  find that $z_{da}\leq1$. Finally, we reproduce an effective $f(z)$ and  show that this class of models is fairly well compatible with present-time data. To do so, we get numerical constraints employing  Monte Carlo fits with the Union 2.1 supernova survey and with the Hubble measurement data set.
\end{abstract}


\begin{multicols}{2}


\section{Introduction}

The $\Lambda$CDM model predicts the existence of a cosmological constant $\Lambda$, entering the Einstein equations and responsible for the dynamical properties of the universe today. Indeed, the model arguably represents the simplest cosmological paradigm to allow the universe to speed up at late times. Despite its experimental successes, the $\Lambda$CDM model is unable to explain why the magnitudes of both matter and $\Lambda$ energy densities are so comparable today. Nevertheless, cosmological constraints indicate a severe difference between the observed $\Lambda$ and the corresponding quantum field theory's value. It follows that the standard $\Lambda$CDM model could represent only a limiting case of more general paradigm, based on the existence of some sort of (dynamical) cosmic fluid, driving the universe today, dubbed dark energy \cite{darkenergy}. Dark energy seems to dominate over all cosmological species enabling the universe to accelerate. Dark energy acts as a \emph{dynamical source} for speeding up the universe, whereas dark matter is responsible for structure formation. Recently, the need of understanding at which time the decelerating-accelerating transition phase occurs has leads cosmologists to directly measure the corresponding transition redshift $z_{da}$ \cite{ratra13,other}. There exists a wide consensus, based on robust observational supports, indicating $z_{da}$ around the unity \cite{ratra13,alt}. Even though the $\Lambda$CDM value of $z_{da}$ reproduces the same result of recent observations, we cannot conclude that dark energy is a constant at all stages of universe's evolution. However, several classes of models depart from the case of $z_{da}\sim1$, indicating that $z_{da}$ represents a {\emph new cosmic number} which any cosmological model should reproduce to be in accordance with cosmic bounds. 
Among concurring models to $\Lambda$CDM, modified gravity is considered  a  realistic alternative to dark energy.  In particular, the challenge of finding out the most viable function $f(R)$, which permits to describe the gravity as a whole, passes through the reconstruction of the correct cosmological model. In other words, determining viable numerical outcomes of $z_{da}$ means to reproduce the $f(R)$ models which satisfy the late-time bounds in a model independent way \cite{nos}. To support  this idea, one can  consider  \emph{ cosmography}, which represents a technique for matching cosmic data with observable quantities without imposing  a particular cosmological model. In general,  a Taylor expansions evaluated at present-time is adopted in cosmography.  Constraining a class of coefficients gives rise to  the \emph{cosmographic series} \cite{luongo}. Hence, the use of cosmography may suggest possible departures from the standard $\Lambda$CDM model and also allows to understand how and whether dark energy evolves in time \cite{due}. In this paper, we show how to get the  Hubble rate by assuming the cosmographic constraints on $f(R)$ models. Further, we bound numerical intervals for the transition redshift $z_{da}$ and  demonstrate that only particular  classes of $f(R)$ models allow to get $z_{da}$ in the correct observable ranges. Afterwards, we propose an effective evolving dark energy term which extends the $\Lambda$CDM model and provides a logarithmic correction to the dark energy density. This model predicts that the corresponding acceleration parameter changes sign around $z\leq1$. Finally, we describe the corresponding $f(z)\equiv f(R(z))$ function in terms of the cosmographic coefficients and  propose a possible expression for it. The paper is structured as follows: in Sec. II we describe the cosmographic approach  in the context of $f(R)$ gravity and  compute the corresponding transition redshift. We also show a comparison with current data and  demonstrate that the dark energy corrections are compatible with  observations. We  also report  numerical outcomes. Finally, Sec. III is devoted to conclusions and perspectives.

\begin{table*}
\caption{{\small Best fits performed by considering Monte Carlo techniques based on the Metropolis algorithm evaluated with 1$\sigma$ error bars for each parameter. We perform a fit by using supernovae without imposing any priors, i.e. adopting a free experimental test. We also use $H(z)$ data without imposing any assumptions. The experimental tests have been carried out by using the luminosity distance and the definition of $H$ respectively. The transition redshifts here reported have been found using the definition of $z_{da}$ for the model, whereas error bars through standard logarithmic propagation. All is adimensional except  $H_0$ which is reported in units of km s$^{-1}$ Mpc$^{-1}$.}}

\begin{tabular}{c c c c}
\hline
\hline

fit 			& supernovae &  $H(z)$ \\ [0.8ex]
\hline
$p$ value		&  {\small $0.692$ }
                &  {\small $0.960$ }\\[0.8ex]

\hline
{\small$ H_0$}		& ~{\small $68.070$}{\tiny${}_{-2.200}^{+3.330}$}~	
                            & ~{\small $68.490$}{\tiny${}_{-2.500}^{+3.390}$}~\\[0.8ex]

{\small$ q_0$}		& ~{\small $-0.542$}{\tiny${}_{-0.083}^{+0.072}$}~	
                            & ~{\small $-2.941$}{\tiny${}_{-0.043}^{+0.092}$}~\\[0.8ex]

{\small$ j_0$}		& ~{\small $0.577$}{\tiny${}_{-0.353}^{+0.448}$}~	
                            & ~{\small $-0.955$}{\tiny${}_{-0.175}^{+0.228}$}~\\[0.8ex]

{\small$\Omega_m$}		& ~{\small $0.214$}{\tiny${}_{-0.041}^{+0.039}$}~	
                        & ~{\small $0.272$}{\tiny${}_{-0.033}^{+0.034}$}~\\[0.8ex]

{\small$\alpha$}		& ~{\small $2.301$}{\tiny${}_{-0.169}^{+0.165}$}~	
                        & ~{\small $1.966$}{\tiny${}_{-0.199}^{+0.231}$}~\\[0.8ex]

{\small$\beta$}		& ~{\small $0.760$}{\tiny${}_{-0.471}^{+0.643}$}~	
           			& ~{\small $0.285$}{\tiny${}_{-0.537}^{+0.773}$}~\\[0.8ex]
		
{\small$z_{\mathrm{da}}$}		& ~{\small $0.860$}{\tiny${}_{-0.272}^{+0.289}$}~	
                            & ~{\small $0.632$}{\tiny${}_{-0.140}^{+0.161}$}~\\[0.8ex]

\hline \hline

\end{tabular}

\label{tab:dL2}
\end{table*}

\section{Cosmographic reconstruction of  transition redshift in $f(R)$ models}

Corrections to the Hubble rate, inferred from $f(R)$ gravity, can be achieved  by means of cosmography. The technique allows  to  determine the corresponding $f(R)$-transition redshift $z_{da}$ showing its compatibility with present-time constraints. In particular, cosmography represents a technique to fix limits on $f(R)$ and its derivatives  by passing through the use of an auxiliary function  $f(z)$. This means to  define $f(z)\equiv f(R(z))$ and its first derivative through ${df(R(z))\over dR}\equiv\frac{df(z)}{dz}\left(\frac{dR}{dz}\right)^{-1}$. To this end, one needs the scale factor expansion into a Taylor series around the present epoch $t_0$:
\begin{equation}\label{aexpans}
a(t)=\sum_{m=1}^{\infty}\frac{d^m a}{dt^m}\Big|_{t-t_0=0}\frac{(t-t_0)^m}{m!}\,.
\end{equation}
Keeping in mind Eq. (\ref{aexpans}), we are able to evaluate $f(z)$ and derivatives at $z=0$ which corresponds to $t=t_0$. To relate $f(z)$ and derivatives to the cosmographic expansion of $a(t)$, it is necessary to assume the constraint on $R$ \cite{capoz1}
\begin{equation}\label{eq: constr}
{R\over 6H} = (1+z)\ {H}_{z} - 2 {H}\,,
\end{equation}
with $H_z\equiv\frac{dH}{dz}$. Simple algebra leads to $R_0 =\,  6{H}_0\left({H}_{z0}-2{H}_0\right)$ and $R_{z0} =\, 6{H}_{z0}^2-{H}_0(3{H}_{z0}-{H}_{2z0})$. From Eq. (\ref{aexpans}), we define the cosmographic coefficients as
\begin{eqnarray}
     &H(t)& = \frac{1}{a}\frac{da}{dt}\,, \\
    &q(t)& = -\frac{1}{a     H^2} \frac{d^2a}{dt^2}=-\left(1+\frac{\dot H}{H^2}\right)\,, \\
    &j(t)& = \frac{1}{a     H^3} \frac{d^3a}{dt^3} = -(3q+2)+\frac{1}{{H}^3}\frac{d^2 {H}}{dt^2}\,,
\end{eqnarray}
respectively the Hubble rate $H$, the acceleration parameter $q$, the variation of acceleration, i.e. the jerk parameter $j$. Constraining the above quantities by means of cosmological data permits one to fix the initial settings on the curvature dark energy term and so on the form of $f(R)$. Moreover, once scalar curvature is somehow fixed by geometrical observations, cosmography becomes a pure model-independent treatment to fix cosmic bounds and to understand which corrections due to $f(R)$ models are really viable to describe current  universe expansion. Furthermore, it is easy to show that the measurements of the cosmographic parameters provide the disadvantage that all coefficients $H,q$ and $j$ cannot be measured alone. Assuming, in fact, the luminosity distance as function of the cosmographic series, i.e. $D_L=D_L(H_0,q_0,j_0)$, and comparing cosmic data with it, we notice that it is impossible to \emph{decouple} the terms $H_0^{-1}(1-q_0)$ and $H_0^{-1}(3q_0^2+q_0-j_0-1)$, respectively the second and third orders of the Taylor expansion of $D_L$ \cite{cosmoaltro}. Thus, to better limit $H_0,q_0$ and $j_0$ independently, alleviating the corresponding degeneracy problem between coefficients, and also reducing the error bars, one can measure $H_0$ alone, by considering the first order of  $D_L$ Taylor expansion 
\begin{equation}\label{H0}
D_{L} \sim H_{0}^{-1}z\,,
\end{equation}
in the observational range $z\leq0.3$. Here, we adopt the Union 2.1 supernova compilation \cite{union21} and we compare our results with those predicted by the Planck mission \cite{planck}. Once $H_0$ is fixed, it is possible to determine the corresponding $q_0$ and $j_0$ and it is also possible to obtain
\begin{eqnarray}
{H_{z0}\over H_0}=&1+q_0\,,&\label{Hinz0}\\
{H_{2z0}\over H_0}= &j_0-q_0^2\,.&\label{Hinz0bis}
\end{eqnarray}
Finally, we have
\begin{eqnarray}
q_0=&{f_0\over 2H_0^2}+2\,,&\label{etuno} \\
j_0=&{f_{z0}\over 6H_0^2}+q_0+2\,.&\label{etdue}
\end{eqnarray}
These expressions are needful for setting priors on $f(z)$ in terms of cosmography and to understand which deceleration-acceleration transition redshift $z_{da}$ is predicted in the context of $f(R)$ gravity. With these considerations in mind, we can integrate the modified Friedmann equations, directly by using cosmographic measurements as a sort of \emph{setting conditions}. In so doing, we infer a parameterized cosmological model, which differs from the $\Lambda$CDM model and shows an effective logarithmic dark energy term $\Omega_X$ of the form \cite{nos}:
\begin{equation}\label{kjdgh}
\Omega_X=\log(\alpha+\beta z)\,.
\end{equation}
The whole Hubble rate becomes ${H}(z)={H}_0\sqrt{\Omega_m(1+z)^3+\Omega_X}$, with $\alpha$ and $\beta$ constants to be fixed by present-time bounds. In fact, to get ${H}={H}_0$,  terms $\alpha$ and $\beta$ are
\begin{eqnarray}\label{lkdfj}
\alpha&=&\exp(1-\Omega_m)\,,\\
\beta&\in& [0.01,\,\,  0.1]\,,
\end{eqnarray}
which are consistent with cosmographic bounds. The corresponding cosmographic $f(z)$ function becomes
\begin{equation}\label{jhd}
f(z)\approx \tilde f_0+\tilde f_1\, a^{-\sigma_1}+\tilde f_2\,a^{-\sigma_2}+a\,,
\end{equation}
with $\tilde f_0\sim-10$, $\tilde f_1\sim 7$, $\tilde f_2\sim-3.7$, $\sigma_1=1$ and $\sigma_2=2$ and $a\equiv(1+z)^{-1}$. This $f(z)$ predicts the above modified Hubble rate and provides a deceleration-acceleration transition redshift of the form \cite{nos}
\begin{equation}\label{quxz}
q_{f(R)}=-1 + \frac{(1+z)}{2}\frac{\left[3\Omega_m(1+z)^2+{\beta}(\alpha + \beta z)^{-1}\right]}{ \left[\Omega_m(1+z)^3+ \ln(\alpha+\beta z)\right]}\,.
\end{equation}
By means of Eqs. (\ref{lkdfj}) and (\ref{quxz}), the  cosmographic reconstruction favors values of the Hubble constant in agreement with other previous estimations \cite{ar1,ar2,ar3} and with the Planck results on $H_0$ \cite{planck}. The bounds on $f_0$ and $f_{z0}$ suggest that $q$ changes sign at $z_{\mathrm da}\leq 1$, in agreement with previous measurements \cite{ratra13}, showing meanwhile that our model provides $z_\mathrm{da}$ to be compatible with the $\Lambda$CDM predictions. In addition, our cosmological model, providing an evolving dark energy, as presented in Eq. (\ref{kjdgh}) estimates viable $\Omega_m$ and $\beta$, with $z_\mathrm{da}\in[0.57,0.97]$, providing slight departures from the standard $\Lambda$CDM model. Further, our model seems to pass experimental tests, performed using Monte Carlo analyses and the union 2.1 and Hubble rate measurement data sets \cite{ratra13,union21,ohd}
All our numerical outcomes have been listed in Table I. Future improvements will better distinguish any significant deviations from the $\Lambda$CDM case at higher redshift domains. With our results, in fact, it seems that dark energy slightly evolves also at small redshift, although a definitive conclusion remains object of future developments.


\section{Concluding remarks}

The passage between the deceleration to acceleration phases happens as the acceleration parameter $q$ vanishes, corresponding to a precise transition redshift $z_{da}$. To understand how well $f(R)$ gravities are able to reproduce suitable bounds on $z_{da}$, we proposed to reconstruct the class of $f(R)$ functions by means of cosmography. Cosmography represents a strategy to fix constraints on the principal observable quantities by means of Taylor expansions around present-time. Cosmic data may be directly matched with the cosmographic coefficients and so the cosmographic series permits to understand which models are really favored than others. In the case of $f(R)$ gravity, we defined the auxiliary function $f(z)$ as $f(z)\equiv f(R(z))$ and its first derivative as $\frac{df(R)}{dR}=\frac{df(z)}{dz}\left(\frac{dR}{dz}\right)^{-1}$. Thus, we showed how to relate them to the cosmographic parameters, i.e. $H_0, q_0$ and $j_0$. In particular, reconstructing the modified dark energy term in terms of $f(R)$, by using as \emph{initial settings} the numerical values inferred from cosmography, we found that dark energy seems to evolve as a curvature fluid proportional to a logarithm, at small redshift. This fact departures from the cosmological standard model, in which dark energy is a constant at all stages of universe's evolution. However, albeit our model showed an evolving dark energy term, its transition redshift $z_{da}$ is compatible with present-time observations. We also found the corresponding $f(z)\approx \tilde f_0+\tilde f_1\, a^{-\sigma_1}+\tilde f_2\,a^{-\sigma_2}+a$. To find out $f(z)$,  the degeneracy between cosmographic coefficients are alleviated, fixing $H_0$, by  fitting the first order Taylor expansion of the luminosity distance in the range $z\leq 0.3$. All  numerical results are in agreement with present-time observations and it seems possible that dark energy may be framed in terms of a curvature dark energy fluid. To show that  paradigm is capable of well reproducing cosmic bounds, we employed experimental tests, making use of the most recent Union 2.1 supernova survey and of the Hubble measurement data set. Future developments will assume different redshift domains to better fix the cosmographic coefficients. Thus, the corresponding effective dark energy term would change correspondingly, providing a different transition redshift. This technique will be useful to understand whether $f(R)$ functions are compatible with cosmological data at different epochs. Moreover, another significative aspect will be to directly relate the cosmographic coefficients to the transition redshift itself, opening the possibility to use $z_{da}$ as a \emph{direct} cosmographic coefficient.

\end{multicols}

\end{document}